\documentclass[12pt]{article}

\usepackage{amsmath,amsthm,amssymb}

\newcommand{\causet}{\mathcal{P}}
\newcommand{\chn}{C}
\newcommand{\chns}{\mathcal{C}}
\newcommand{\dff}{\sc}

\newcommand{\eg}{\emph{e.g.},\,}
\newcommand{\etal}{\emph{et al.}\,}
\newcommand{\ie}{\emph{i.e.},\,}

\newcommand{\mm}{\mathbf{M}}
\newcommand{\mmm}{\mathcal{M}}

\title{Glafka 2004: Spacetime topology from the tomographic histories
approach I: Non-relativistic Case}

\author{Ioannis Raptis\footnote{EU Marie Curie Reintegration
Postdoctoral Research Fellow, Algebra and Geometry Section,
Department of Mathematics, University of Athens,
Panepistimioupolis, Athens 157 84, Greece; {\em and} Visiting
Researcher, Theoretical Physics Group, Blackett Laboratory,
Imperial College of Science, Technology and Medicine, Prince
Consort Road, South Kensington, London SW7 2BZ, UK; e-mail:
i.raptis@ic.ac.uk}, Petros Wallden\footnote{Theoretical Physics
Group, Blackett Laboratory, Imperial College of Science,
Technology and Medicine, Prince Consort Road, South Kensington,
London SW7 2BZ, UK; e-mail: petros.wallden@imperial.ac.uk}  and
Rom\`an R. Zapatrin\footnote{Department of Information Science,
The State Russian Museum, Inzenernaya 4, 191186, St. Petersburg,
Russia; e-mail: zapatrin@rusmuseum.ru}}

\date{}

\begin{document}

 \maketitle

\begin{abstract}
The tomographic histories approach is presented. As an inverse
problem, we recover in an operational way the effective topology
of the extended configuration space of a system. This means that
from a series of experiments we get a set of points corresponding
to events. The difference between effective and actual topology is
drawn. We deduce the topology of the extended configuration space
of a non-relativistic system, using certain concepts from the
consistent histories approach to Quantum Mechanics, such as the
notion of a record. A few remarks about the case of a relativistic
system, preparing the ground for a forthcoming paper sequel to
this, are made in the end.
\end{abstract}

\section{Introduction with Motivational Remarks}

In the standard formulation of relativity theory, the spacetime
topology is {\it a priori} fixed by the theorist to that of a
continuous manifold; hence, it is not an observable entity. Only
the metric structure is traditionally supposed to be dynamically
variable. With the exception of Wheeler's celebrated, but largely
heuristic, spacetime foam scenario \cite{wheel}, there is no well
developed theory in which the spacetime topology can be regarded
as a dynamical variable proper, with quantum traits built into the
theory from the very start. However, one may try to consider
idealized situations where certain topological features are
represented as quantum variables that can in principle be observed
and measured \cite{Zaptop}. Even in General Relativity (GR), where
no variable quantity is supposed to be quantum---\ie subject to
coherent quantum superpositions and associated uncertainty in its
determinations, we need histories (\eg material particles' causal
geodesic trajectories) to actually define the topology of
spacetime. This is because the concept of neighborhood turns out
to be something which someone (:an observer), located at some
point in spacetime, deduces for regions that belong to her causal
past. Similarly, the concept of distance can be established only
if information (:causal signals, or actual travelling material
particles) is (causally) transmitted from one point to another.
All in all, the causal nexus of the world determines both its
topological and metric structures.

On the other hand, an interesting feature of quantum mechanics is
that we may be able to make and verify statements about topology
from a single-time case, as long as we are allowed to repeat
experiments (and in principle we are allowed to do that
indefinitely, if only in a theoretical, idealized,
`gedanken'/theoretical fashion) in order to get the relative
frequencies. In the classical (\ie non-quantum mechanical) case,
one-time measurements do not give any information about global
properties, such as the background topology.

Having said that, a remarkable consequence of quantum mechanics is
that the wavefunction is a non-local entity, so that we may be in
a position to deduce topological properties of the background,
provided that we have enough repetitions of the experiment to
reconstruct the relative frequencies. Thus, instead of saying that
the wavefunction is a square integrable function on a topological
space and use this to deduce probabilities about experimental
outcomes (:events), we hereby propose to do the converse. \emph{We
start from probabilities and the continuity assumption for events,
and from this information we derive the structure of the
topological space in which these events are supposed to happen}. A
word of caution is due here: the continuity assumption is normally
taken to presuppose a topology---for how else can one talk about a
\emph{continuous} wavefunction? Well, and here is the crux of the
inverse scenario: our assumption is that the wavefunction must be
continuous with respect to the topology \emph{to be deduced} from
the relative frequencies of events. In other words, \emph{the
(continuity of the) wavefuction is born with the topology being
deduced}.

In what follows, we do the same for the 4-dimensional case and
recover `spacetime'.  We should point out that no matter that we
talk about space-time, we are still in the non-relativistic
regime. We just speak of space points labelled by their
`absolute', Galilean time of occurrence. The relativistic case
will be considered in a forthcoming paper \cite{case II}.

One could say that histories are still needed to define global
properties, such as the topology; however, here we maintain
instead that they are needed in order to extract the {\em form} of
the wavefunction. Of course, {\it prima facie} one can counter our
arguments by holding that the assumption about (complete)
knowledge of the wavefunction immediately leads to EPR-type of
paradoxes. Our retort is that EPR-phenomena do not arise in our
setting, while causality is rescued by the fact that we need
classical communication to recover the full (:complete) state, as
for example in various (quantum) teleportation scenarios---see,
\eg Aharonov and Vaidman \cite{ahar}.

Let us outline the contents of the present paper. In the first
part we introduce the consistent histories approach whereby we are
given a configuration space for the system, its full Hamiltonian
(including interactions), as well as the initial conditions
(generally speaking, `exosystemic' parameters of the problem
traditionally supposed to be determined by an experimenter
external to the experimentee---the physical system under
experimental focus), and from these we calculate the probabilities
for histories to occur. In our \emph{inverse}---{\it alias},
`\emph{tomographic}'---approach, we are given the sets of observed
histories together with their relative frequencies, and from these
we reconstruct (some of) the parameters of the problem, with no
allusion to external/internal systemic distinctions, as befits the
histories approach. Then, certain issues about topology and the
character of various possible indeterminacies of the derived
topology that are involved in our approach are highlighted.

The main part of the paper follows, where we present \emph{what}
we are able to recover and \emph{how} we do that. In this paper we
specifically develop the \emph{non-relativistic} case and focus on
what can be said about topology using the set of histories alone,
and also what needs some further measurements in order to be
`sharply' determined. Finally, we illustrate all this by virtue of
two toy-models. The first is our variant of the usual double-slit
experiment, both when the particle is detected at the slit, and
when it is not. The second is an example of an environment
involving a `bath of sensors'.

But before we delve into the paper, we feel that the new term
`\emph{tomography}' ought to be further explained; otherwise,
there is no reason to have it only for the sake of fancy
neologisms and lexiplacy. We believe that its use can be justified
on the following semantic grounds: experiments and their records
may be thought of as `cuts' (:`$\tau o
\mu\acute{\epsilon}\varsigma$ in Greek) incurred on the quantum
system.\footnote{Recall the Heisenberg `schnitts' (German for
`cuts') in the standard Copenhagean quantum theory.} From (the
results of) these `observational measurement-slices' and their
relative frequencies of occurrence, we `retro-write' (:`redraw',
or `reconstruct retrodictorily' so to speak)---as it were, `after
the fact'---the (spacetime) topology. Moreover, in Greek, the verb
`to write' (or `to draw', generically speaking) is
$\gamma\rho\acute{\alpha}\phi\omega$. Hence
`\emph{tomo-graphy}':\footnote{In Greek, `$\tau o\mu
o$-$\gamma\rho\alpha\phi\acute{\iota}\alpha$:=`slice-wise
writing/skethching/drawing').} we are re(tro)sketching spacetime
topology from `experimental cuts' exercised on the quantum
system(!) All in all, this etymological dissection of the word
`tomography' accords with the title of the paper:
``\emph{spacetime topology (derived, or effectively re-sketched)
from `tomographic', inverse histories}''.

\section{Histories and inverse histories approach}\label{shist}

The decoherent histories approach to quantum mechanics deals with
the kind of questions that may be asked about a closed system,
without the assumption of wavefunction collapse (upon
measurement). It tells us, in a non-instrumentalist way, under
what conditions we may meaningfully talk about statements
concerning histories of our system, by using ordinary logic. This
approach was mainly developed by Gell-Mann and Hartle
\cite{GH90a,GH90b,GH90c,Har91a,Har91b,GH92,Har93a}, and it was
largely inspired by the original work of Griffiths \cite{Gri84}
and Omn\`es \cite{Omn88a,Omn88b,Omn88c,Omn89,Omn90,Omn92}.

In this section, after we briefly recall useful rudiments of the
standard histories scheme, we introduce its `converse' theoretical
scenario that interests us presently: \emph{the inverse histories
approach}. Pictorially, the two schemes are related as follows:

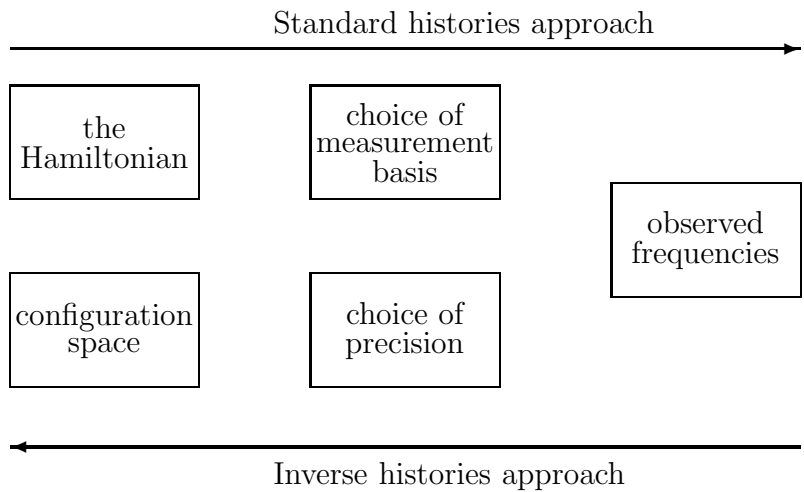
\unitlength1mm
\[
\begin{picture}(120,70)
\thinlines \put(0,40){\framebox(25,15){\shortstack{the\\
Hamiltonian}}}
\put(0,15){\framebox(25,15){\shortstack{configuration\\ space}}}
\put(40,40){\framebox(25,15){\shortstack{choice of\\ measurement\\
basis}}} \put(40,15){\framebox(25,15){\shortstack{choice of\\
precision}}} \put(80,27){\framebox(25,15){\shortstack{observed\\
frequencies}}} \thicklines \put(0,60){\vector(1,0){105}}
\put(105,7){\vector(-1,0){105}} \thinlines
\put(35,62){\mbox{Standard histories approach}}
\put(35,2){\mbox{Inverse histories approach}}
\end{picture}
\]

\subsection{The HPO version of the standard histories approach}\label{shpo}

The formulation of the standard histories scenario that we follow
presently is due to Isham \etal (\eg see
\cite{Ish94,Isham:1994uv}), and it is called HPO (History
Projection Operator) approach. It consists of a space of histories
$\mathcal{UP}$, which is the space of all possible histories of
the closed system in question, and a space of decoherence
functionals $\mathcal{D}$. Parenthetically, the space of histories
is usually assumed to be a tensor product of copies of the
standard QM Hilbert space. Two histories are called disjoint,
write $\alpha\perp\beta$, if the realization of the one excludes
the other. Two disjoint histories can be combined to form a third
one $\gamma=\alpha\vee\beta$ (for $\alpha\perp\beta$). A complete
set of histories is a set $\{\alpha_i\}$ such that
$\alpha_i\perp\alpha_j\quad(\forall\alpha_i,\alpha_j,\quad i\neq
j)$, and
$\alpha_1\vee\alpha_2\vee\ldots\vee\alpha_i\ldots=\mathbf{1}$

A decoherence functional is a complex valued function
$d:\mathcal{UP}\times\mathcal{UP}\longrightarrow\mathbb{C}$ with
the following properties:
\begin{itemize}\item[a)]Hermiticity: $d(\alpha,\beta)=d^*(\beta,\alpha)$
\item[b)]Normalization: $d(1,1)=1$ \item[c)]Positivity:
$d(\alpha,\alpha)\geq0$ \item[d)]Additivity:
$d(\alpha,\beta\oplus\gamma)= d(\alpha,\beta)+d(\alpha,\gamma)$
for any $\beta\perp\gamma$

\end{itemize}

A complete set of histories $\{\alpha_i\}$ is said to obey the
{\dff decoherence} condition, \ie
$d(\alpha_i,\alpha_j)=\delta_{ij}p\left(\alpha_i\right)$ while
$p\left(\alpha_i\right)$ is interpreted as the probability for
that history to occur \emph{within the context of this complete
set}.

The decoherence functional encodes the initial condition as well
as the evolution of the system. Here we should also note that the
topology of the space-time is presupposed when we group histories
into complete sets, \ie in collections of partitions of unity.

In standard QM, histories correspond to time ordered strings of
projections and to combination of these when they are disjoint. An
important issue here is the relation between decoherence and
records. Namely, it can be shown that if a set of histories
decoheres, there exists a set of projection operators on the final
time that are perfectly correlated with these histories and vice
versa.\footnote{This is the case for a \emph{pure} initial state,
and we restrict ourselves to it.} These projections are called
{\em records}. It is this concept that figures mainly in our
approach (\eg see Halliwell \cite{Hal99}).

To sum things up, in the standard histories approach
\begin{itemize}
\item The system is given, as well as its environment. The latter
is represented by prescribing initial conditions and in some cases
final conditions.

 \item The space, its topological structure
in particular, is presupposed.

 \item The interactions are
given in terms of the decoherence functional, which encodes the
dynamical information. For the complete dynamics, the full
Hamiltonian must be known.
\end{itemize}

\subsection{Tomographic histories approach}\label{sinvhist}

In our approach things are different, as we solve the
\emph{inverse} problem. While in standard histories one is given
the Hamiltonian, initial conditions, as well as the space on which
they are defined, and the aim is to predict probabilities for
histories, we do the opposite thing. We make repetitions to get
the frequencies for different records. Then, by making certain
assumptions about these records, namely, that they are nothing but
records of {\em events}, we recover the topological structure of
the underlying configuration space. This means that from a set of
events, with no other structure presupposed (:{\it a priori}
imposed from outside the system), we end up with a causal set
representing the discretized version of the \emph{extended
configuration space} of the system in question.

The extended configuration space that we get will be an
`\emph{effective}' one, and in a sense it accounts for certain
properties of the Hamiltonian, such as interactions with other
objects not controlled by the experimenter. For instance, the
latter could be some kind of `repulsive' field that prohibits the
system to go somewhere (:in a region of its configuration space),
which can then be recovered as a hole (:a dynamically inaccessible
region) in that space.

\medskip

To compare the two approaches, let us review for a moment the
standard histories approach where the decoherence functional, as
well as the space of histories, are given. For these, one is
assumed to be given the initial conditions, the configuration
space, and the Hamiltonian of the system in focus---\ie generally
speaking, the parameters of the system. When we are able to
perform multiple runs of the experiment and we choose a decohering
set of histories, the decoherence functional yields the
probability for each history to occur, which, in turn, corresponds
to the history's relative frequency with respect to the set
chosen.

Having the same Hamiltonian and the same initial conditions, we
may consider another decohering set of histories, not necessarily
compatible with the previous one, for which again the
probabilities can be calculated. This is in broad terms what the
usual histories approach accomplishes.

\medskip

We on the other hand will be tackling the inverse problem. The
essence of our approach is the following. Since we can carry out
our experiment sufficiently many times, we have access to the
following two things---the set of possible histories and the
relative frequencies for each history to occur for every initial
state. From this we recover the parameters of the experiment,
namely, the effective topology of the extended configuration
space.

One thing to highlight here is what corresponds to a decoherent
set of histories in our inverse scenario. It is one particular
partition of unity of the record space. Our freedom of choosing a
particular basis in which to measure things will in general give
different decoherent sets than had we chosen a different one
(:different basis, different decoherent sets). Note also that
since we consider histories \emph{operationalistically}, we always
deal with histories that are contained in a decoherent set,
namely, the set that corresponds to the set of records that we
choose to analyze.

In our setup we shall assume that \emph{the records capture the
spatio-temporal properties} (of the system in focus). This means
that the histories are coarse-grained trajectories of the system,
belonging to a space whose topological properties we ultimately
wish to deduce. We shall then claim that the whole concept of
spacetime, as a background structure, does not make sense in
finer-grained situations. In this way, all the histories are
single-valued on our discretized version of `effective spacetime'.
One should note here that we may still have histories that have
the particle in a superposition of different position eigenstates,
but only if the latter are `finer' than the degree of our
coarse-graining. With the coarse-graining we effectively identify
(\ie we group into an `equivalence class' of some sort) the points
that we cannot distinguish operationally, with the resulting
equivalence class of `\emph{operationally indistinguishable
points}' corresponding to a `blown up', `fat point' in our
discretized version of `effective spacetime'.

\subsection{Classical versus quantum indeterminacy of topology}\label{scqtop}

In this subsection we would like to emphasize that there are two
essentially different kinds of indeterminacy involved in
derivations of the effective topology.

The first one is of a `classical' character, that is, it comes
from the lack of \emph{our} knowledge about the systems'
configuration space, as for instance when we do not have
sufficiently many repetitions of the experiment. For example, the
configuration space might appear to be a segment of a straight
line, when in fact it is a circle. This could be due to incomplete
information that we gather from an insufficiently repeated
experiment, which could result to some points at the end of the
segment, that would ultimately make the configuration space a
circle, not to be detected. Another way that classical
indeterminacy could arise would be when some records are simply
not accessible\footnote{This is not the case in this paper. In our
setup we assume that we have access to all records that are
related to detectable events.} when, as a matter of fact, the
interaction of our system with the `record space' is supposed to
capture all the spatiotemporal features or properties of the
system. {\it In toto}, as befits the epithet `classical', this
type of indeterminacy in effective topology determinations is an
`\emph{epistemic}' one: it reflects \emph{our} ignorance, our
partial experimental knowledge about and control over the quantum
system.

The second type of indeterminacy, like the one arising in Quantum
Theory, is due to a fundamental `quantum dichotomy' of our
experimental settings and determinations. For instance, the
topology of coordinate and momentum space of a quantum particle
may be different from each other, so that what we recover in the
end depends on what we initially choose to measure: coordinates or
momenta. Plainly, this reflects the fundamental quantum duality
between the position and momentum observables in standard QM,
which in turn is a reflection of the \emph{ontological} (as
opposed to epistemic) nature of quantum indeterminacy and
uncertainty. Our setup simply limits our freedom to measure
anything we want to what is produced by a decoherent set of
histories, and therefore it is associated with a projection
operator on our record space. We must emphasize however that we
still have some freedom, since incompatible consistent sets have
incompatible records in the record space, so that our choice of
what basis to measure in the record space is still in force. This
issue is addressed in more detail in section \ref{sstatrectop}.

\subsection{The operationalistic underpinnings of our scenario}\label{sopersetup}

Our approach is essentially \emph{operationalistic}. The notion of
\emph{record space} is regarded as the only source of information
we possess about the system we wish to explore. The effective
topology then refers to the configuration space of the system in
question. In our tomographic approach, we are given the sets of
observed histories together with their relative frequencies, from
which then we reconstruct the parameters of the problem.

We assume that some of the records may be identified with
particular events, \ie spacetime `points'. Furthermore, we claim
that this is the only case we may speak of a configuration space
proper. That is, if we do \emph{not} have access to events even in
principle, we \emph{cannot} speak about their support or their
topological and causal nexus, as, say, in the causal set scenario
(causet). Then, relative frequencies are recovered by repetition
of the whole histories involved: by restarting the system in an
identical environment and letting it evolve for the same amount of
time.\footnote{From our vantage, `\emph{history could in principle
repeat itself}' (pun intended).} In our operationalistic
(ultimately, relational-algebraic) view, the only way one can talk
about some background structure such as `spacetime', is relative
to something else. More precisely, we use our data (records) to
(re)construct an `arena' for a particular subsystem of the
universe that we are interested in, and it is \emph{only} in this
sense that we may speak of `spacetime'. Retrodictorily,
\emph{`spacetime' is where and when `it' must have happened, if we
judge by our records, and the latter are the only data we have
got}. Thus, philologically speaking, `\emph{quantum tomography is
spacetime archaeology}'.

More precisely, we have a system (call it `particle'), which is
placed into an appropriate experimental environment, and we are
able to

\begin{itemize}

\item Repeat the experiment with \emph{the same} initial
conditions. In this way we get the relative frequencies of the
records.

\item Vary the initial conditions of the system in question,
leaving all the environment (and records) the same. For each
initial condition of the system, we rerun the experiment. These
first two steps give us the set of all possible histories
(coarse-grained trajectories) of the particle, as well as their
relative frequencies.
\end{itemize}

\noindent Another basic ingredient is the space of records. It is
a space of data resulting from controlled environment tampering
with the system, and it is supposed to capture its spatiotemporal
properties. Records are interpreted as \emph{distinguishable
events}. That is to say,

\begin{itemize}

\item We can distinguish them spatiotemporally. Although we do not
know the structure of the set of records that corresponds to
events, we can identify each record corresponding to a spacetime
point as being different from the others. Thus, while we know
nothing {\it a priori} about their causal or spatial (topological)
ordering, events can be labelled so that we do not have
identification problems. For instance, we may consider photons of
different frequencies, each frequency mode coming from one point.
In the examples to follow this will become more transparent.

\item We can vary each record corresponding to a particular event
independently. The variation is in some sense small---this may be
effectuated by a `small energy' variation of the record. The
latter is assumed to be small enough not to affect the `topology'
of the records (\ie neighborhoods in the set of records remain the
same). By `topology' we mean a reticular structure associated with
appropriate coarse-graining of a region of the extended
configuration space we explore. The said variations give us the
proximity relations between events.

\end{itemize}

Experiments are carried out repeatedly and multiply. We label the
runs by initial conditions of the system, number of run and
`positions' of events.\footnote{By this we mean whether or not we
varied one record corresponding to an event.} Each run gives us a
history, \ie a sequence of causally related events. Note here that
for the same initial conditions of the system, the different
histories group together to form decoherent sets.

To conclude, from our experiments we get the following
information:

\begin{enumerate}

\item The set of histories of the system associated with a fixed
set of initial conditions. We call this set of histories {\dff
fiducial set}. Here we emphasize that these correspond to
coarse-grained `trajectories'.\footnote{The inverted commas are
added to the word `trajectories', since the space on which they
are defined is not presupposed.}  We define the \textbf{set of all
histories} to be $\mathcal{C}$, while each history that is
contained in it is denoted by $C_i$. We therefore obtain the set
$\mathcal{C}$ as well as the set $\causet$ which is the set of all
possible events, or else the set of `spacetime' points\footnote{We
remind here that we just speak of space points labelled by their
Galilean time of occurrence}.

\item The relative frequencies of outcome of these histories
depending on the initial conditions. This is a function $f_j:
\mathcal{C}\rightarrow[0,1]$ which gives the (normalized) relative
frequency of histories for each particular initial condition
(corresponding to $j^{th}$ initial state of the system).

\item The change in the relative frequencies when one event is
varied. This is a function  $f^p_j: \mathcal{C}\rightarrow[0,1]$
which is the \emph{new} relative frequencies when the event $p$
has been varied. This will lead us to the proximity relation
between the points produced by the fiducial set of histories.

\end{enumerate}

It is important to note that \emph{before} we vary the records, we
already have the fiducial set of histories. It provides us the set
on which the topology is imposed.

\section{Non-relativistic case}\label{snonrelcase}

 We reconstruct the \emph{effective} topology of the extended
configuration space. But let us explain what we do in a bit more
detail.

\paragraph{Effective versus `real' topology.} In our approach, we
consider the effective topology which we derive from our
observations. That means the following. Believing in Einstein's
theory, we posit that the physical processes take place in
spacetime, which is a topological space with certain \emph{`real'}
topology. However, there is no way for us to measure this `real'
topology exactly. That is why we are speaking of \emph{effective
topology}---the topology of a model of configuration space which
accords with our experiments and fits their outcomes.

An important issue should be emphasized at this point. Suppose we
have derived a non-trivial topology for the configuration
space---say for instance that it has a defect, such as a hole.
This indicates to us merely that we have non-contractible loops,
nothing more. Why these loops fail to be contractible---due to the
existence of a `real hole', or because of, say, the presence of a
potential barrier---such a question is, as a matter of principle,
not verifiable within our approach.

As a consequence, we may admit transitions between 3-dimensional
surfaces of different number of components (with respect to the
effective topology), without regarding this as being unphysical.

\paragraph{The record space.} As noted before, we rely solely on
operationalistic means to recover the effective topology. In turn,
this means that we are able to control the preparation of the
initial state (see section \ref{sopersetup}) and then read out the
observation which is carried out by a specified device. The state
space of this device we shall call {\dff record space}. Here it
should be pointed out that we assume certain things about this
record space. In the case of the examples in section \ref{toy
models}, we specify the main features of the interaction
Hamiltonian of the system with the record. More generally, we need
only to assume that it captures the spatiotemporal properties of
the system and therefore that it leaves records of events. Other
records, outside our record space, may exist and they specify
other features of the particle, such as its spin or electrical
charge. If some records of the spatiotemporal properties are
elsewhere then we may end up with an incomplete topology
reflecting the classical indeterminacy mentioned earlier.

In closing this subsection we should also stress that since the
device is anyway a quantum system, reading out the records causes
some loss of information about the system in focus. Moreover, our
choice of what to read out may also affect the resulting topology,
which is related to the aforementioned ontological quantum
indeterminacy.

\subsection{Extended configuration space and algebraic considerations}\label{sextconfspace}

We have a classical or quantum physical system, and we observe it
for a period of time $(t_0,t_1)$. If $\mm$ is the configuration
space of the system, then the Cartesian product

\begin{equation}\label{e01}
    \mmm=\mm\times(t_0,t_1)
\end{equation}

\noindent is the \emph{extended configuration space}. Moreover, we
also take into account a more general situation in which the
topology of the configuration space $\mm$ may change in time and
the extended configuration space $\mmm$ is no longer decomposable
into a product like \eqref{e01}.

Assume for a moment that the configuration space is at all times
connected. This is not a trivial statement, as we are talking of
`effective extended configuration space' which in principle allows
for transitions from connected to non-connected subsets in
different moments of time. By considering $\mathcal{C}$, the set
of all histories, we may deduce the spatial slices as the subsets
$S_i$ of points no pair of which is contained in the same history
(trajectory).

\begin{equation}
\forall\; p,q\in S_i \implies \nexists\;
C_j\in\mathcal{C}\mid\quad p,q\in C_j
\end{equation}

Moreover, we regard `maximal' slices as being the `time-slices',
\ie any extension of the spatial surface will move the subset
outside the class of spatial surfaces.

\begin{equation}
 \nexists\; r\in \causet \mid r\cup S_i=S_j
\end{equation}

Note here that the relation indicating that two points do not
belong to the same history is \emph{transitive} in the case we
have only one component. It should also be noted that we cannot
determine the order of the slices merely from the set of histories
(\ie without varying the records), neither can we deduce any other
topological feature within each of these slices. Thankfully, the
latter is not the case in the relativistic situation since the
upper bound in the speed of transmission of information leads to a
notion of proximity in each spatial surface. This will be explored
in a later publication \cite{case II}.

Returning to the general case, in which transitions from connected
to non-connected spaces are allowed, the above procedure will
produce ambiguities. Two `events' could never be contained in the
same history due to the fact that they are in separate connected
components and \emph{not} because they `occur' at the same time.
Trying then to form maximal subsets of $\mathcal{P}$ that are not
contained pairwise to any history, will not lead to a unique
partitioning of the set of `spacetime' points . This is due to the
fact that the property of two points not belonging to the same
history is not transitive anymore.

\paragraph{An example of ambiguity in partitioning:}

\par
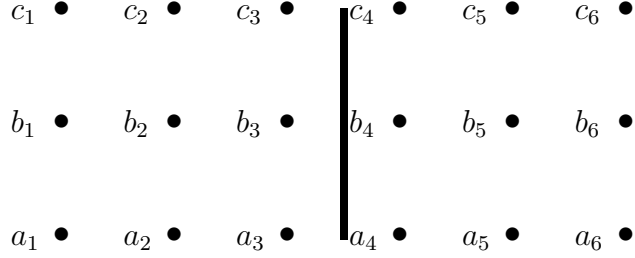
\unitlength1.5mm
\newcounter{palabels}
\newcounter{pblabels}
\newcounter{pclabels}
\begin{center}
\begin{picture}(30,30)

\multiput(-10,0)(10,0){6}{\circle*{1.2}}
\multiput(-10,10)(10,0){6}{\circle*{1.2}}
\multiput(-10,20)(10,0){6}{\circle*{1.2}} \linethickness{.9mm}
\multiput(15,-0.5)(10,0){1}{\line(0,1){20.5}}
\multiput(-14.5,-1)(10,0){6}{$a_{\stepcounter{palabels}
\thepalabels}$}
\multiput(-14.5,9)(10,0){6}{$b_{\stepcounter{pblabels}
\thepblabels}$}
\multiput(-14.5,19)(10,0){6}{$c_{\stepcounter{pclabels}
\thepclabels}$}
\end{picture}
\end{center}

\medskip

We have a non-connected space. Say we have two boxes separated by
a rigid partition (\eg an infinite potential barrier). The thick
line in the graph represents the partition. Apart from the
obstructing partition, all histories are allowed which do not
contain points of the same `horizontal' line corresponding to
`same time'. If the particle is in one time in point $a_1$ at the
left side of the partition, then it can never be in any of the
points on the right hand side of the partition, as \eg point
$b_5$. This, according to the previous definition of `time-slice',
means that $a_1$ is in the same slice with all the points on the
right hand side of the partition no matter which instant they are
measured at.

The latter would lead to contradiction, since clearly $a_5$ and
$b_5$ are not in the same time-slice as there is a history joining
them. If we stick to the proper definition of `time-slice', \ie a
maximal set of points pairwise not belonging to the same history,
we will end up having point $a_1$ in one of the following
`slices': ($a_1,a_2,a_3,a_4,a_5,a_6$), or
($a_1,a_2,a_3,b_4,b_5,b_6$), or, finally
($a_1,a_2,a_3,c_4,c_5,c_6$). Any of these obey the definition of
`spatial-slice'; therefore, just from the set of histories we will
end up with some ambiguity about what a spatial-slice or a `moment
of time' is.

In general, this would not be a problem since we could consider
each component separately. But, in our effective set up we may
have the two disconnected components becoming connected in the
future. For example, the separation was made from ice and it
melted (or from an unstable radioactive substance which quickly
decomposed!). An example of this situation will be examined later.

We could therefore already make one non-trivial statement about
the topology, just by considering the set of histories. Namely,
that if there is a unique way of `foliating' the points of the
`spacetime' into slices, the space is connected. Furthermore, we
will be able to determine the number of different components of
the `4-dimensional' configuration space, and on top of this, the
number of components of one particular `spatial' surface.

\subsection{Extracting connected components}\label{sxtrconnected}

Having the set of decoherent histories, we can already extract
some information about the effective topology. Let us first show
how connected components are detected. In order to do this, recall
that, given a connected component $K$ of a topological space $X$,
the relation $a\sigma b:=\{a,b\in K\}$ is an equivalence relation
on $X$.

\paragraph{4-dimensional connectedness.} In our setup, we are
given the relation $aHb:=\{\exists\chn\in\chns\mid a,b\in\chn\}$,
which means that there exists a history containing both $a$ and
$b$. The relation $\simeq$ is an equivalence, \ie a symmetric,
reflexive and transitive relation on $X$. However, the relation
$H$ is symmetric and reflexive, but not transitive. Thus, the
relation $\sigma$ can be obtained as the transitive closure of the
relation $H$. In general, finding the transitive closure is an
infinite operation; however, here we deal with histories
containing a finite number of events, hence the transitive closure
can be delimited in a finite number of steps.

A possible algorithm to find the transitive closure can be devised
using Boolean matrix machinery \cite{bmql}. Namely, we can define
the relation $H$ by its Boolean matrix (denote it by the same
symbol $H$), then $\sigma$---the transitive closure of $H$---is
obtained as a Boolean matrix power $H^{|A|}$ of $H$, where $|A|$
is the number of antichains. So, effectively the procedure of
extracting connected components goes as follows:

\begin{itemize}

\item Form the Boolean matrix of the relation $H$ `to belong to
the same history'

\[
aHb:=\{\exists\chn\in\chns\mid a,b\in\chn\}
\]

\item Calculate its $|E|$'s power using Boolean arithmetics rather
than $\oplus$ and $\otimes$:

\[
\sigma = H^{|E|}
\]

\noindent Recall that the Boolean operations has the following
rules: $1+0=0+1=1+1=1$, $0+0=0$, $1\cdot 1=1$,
$1\cdot0=0\cdot1=0\cdot0=0$.

\item The resulting matrix $\sigma$ is always block-diagonal and
the blocks of entries are in 1--1 correspondence with the
connected components of the space $E$ of events.

\end{itemize}

\paragraph{Components of a spatial surface.} The procedure just
described would account for the number of components our
`4-dimensional' configuration space has. Note that, since we speak
of `effective configuration space', we may as well have
transitions, in some particular time, from a number of components
to another. It would then be of interest to consider the number of
components a spatial surface has.

To this end we should point out that there is some ambiguity about
what a spatial surface is, thus this ambiguity will also be
present in the considerations to follow.

\begin{itemize}

\item We let $S_i$ be a spatial surface. $\forall\;
p\in\mathcal{P}\backslash S_i\quad$, we consider the following:

$\{S_i^p\subset S_i\mid\forall\; q\in S_i^p\quad\exists\quad
C_j\in\mathcal{C}\mid\quad p,q\in C_j\}$

\item We will then end up with a family of subsets of $S_i$, call
it $\mathcal{S}_{s_i}$. Note that some of these will be identical,
while others may contain others. We declare them `open'.

\item From this family we generate a topology by taking arbitrary
unions and finite intersections of the subsets. The resulting
topology is denoted by $\mathcal{T}_{s_i}$.

\item We then consider a sub-selection of the open subsets of
$\mathcal{T}_{s_i}$ such that:

\begin{enumerate}

\item It covers all $S_i$, \ie their union is $S_i$.

\item They are disjoint.

\item They are `minimal': that is, they contain the smallest of
the subsets in the family $\mathcal{T}_{s_i}$.

\end{enumerate}

This is a disjoint open covering of $S_i$ that is also a basis for
the topology $\mathcal{T}_{s_i}$.

\item Finally, each of these subsets corresponds to one component
of the
 spatial surface $S_i$.

\end{itemize}

\noindent To clarify things, and without wishing to repeat
ourselves, we describe the above in words. We chose the surface in
question. Then, for each point in space we see which part of the
surface is causally connected to it. Then we pick the smallest
family of subsets of the surface that covers the surface. Since
the separate components do not overlap, we need to secure that
this family is also disjoint. That is why we need to generate a
family bigger than $\mathcal{S}_{s_i}$, namely,
$\mathcal{T}_{s_i}$, while from this we are guaranteed to have a
basis that consists of the relevant components, which basis would
{\it a fortiori} be a disjoint covering.

\paragraph{Illustrative example of recovering the components of a spatial surface:}

\par
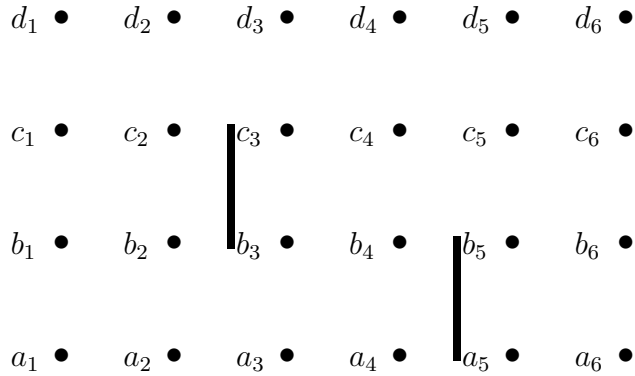
\unitlength1.5mm
\newcounter{alabels}
\newcounter{blabels}
\newcounter{clabels}
\newcounter{dlabels}
\begin{center}
\begin{picture}(30,40)

\multiput(-10,0)(10,0){6}{\circle*{1.2}}
\multiput(-10,10)(10,0){6}{\circle*{1.2}}
\multiput(-10,20)(10,0){6}{\circle*{1.2}}
\multiput(-10,30)(10,0){6}{\circle*{1.2}}
 \linethickness{.9mm}
\multiput(25,-0.5)(10,0){1}{\line(0,1){11}}
\multiput(5,9.5)(10,0){1}{\line(0,1){11}}
\multiput(-14.5,-1)(10,0){6}{$a_{\stepcounter{alabels}
\thealabels}$}
\multiput(-14.5,9)(10,0){6}{$b_{\stepcounter{blabels}
\theblabels}$}
\multiput(-14.5,19)(10,0){6}{$c_{\stepcounter{clabels}
\theclabels}$}
\multiput(-14.5,29)(10,0){6}{$d_{\stepcounter{dlabels}
\thedlabels}$}
\end{picture}
\end{center}

\medskip

Here the space is connected when seen 4-dimensionally. The
partitions that exist forbid for example a history containing
$a_5$ and $b_3$ (the lower one), or $a_5$ and $c_1$ (the higher
partition). Note that even without the `d-column', the space is
connected when viewed `4-dimensionally', as the `transitive
closure' of any point is the set itself.

Now we follow the steps described above. We pick the spatial slice
that corresponds to the b-horizontal line ($b_1,b_2...,b_6$), and
we are looking for its components.

First we consider the set $\mathcal{S}_{s_i}$, which in this case
is the set containing the following subsets:
$\{(b_1,b_2),(b_1,b_2,b_3,b_4),(b_3,b_4,b_5,b_6),(b_5,b_6),(b_1,b_2,b_3,b_4,b_5,b_6)\}$.
Note that the subset $(b_3,b_4)$ does not belong to
$\mathcal{S}_{s_i}$. The result we would like to have is that
there are three components, namely,
$\{(b_1,b_2),(b_3,b_4),(b_5,b_6)\}$. To obtain this, we have to
follow section \ref{sxtrconnected} and extend $\mathcal{S}_{s_i}$
to $\mathcal{T}_{s_i}$, which is the topology induced by
$\mathcal{S}_{s_i}$ if we consider unions and intersections. In
the latter, the subset $(b_3,b_4)$ is also included as it is the
intersection of $(b_1,b_2,b_3,b_4)$ and $(b_3,b_4,b_5,b_6)$.

We then need to pick a sub-selection of the elements of
$\mathcal{T}_{s_i}$ that is disjoint and covers the surface (\ie
the horizontal $b$). There are two possible choices: either
$\{(b_1,b_2),(b_3,b_4),(b_5,b_6)\}$, or
$\{(b_1,b_2,b_3,b_4,b_5,b_6)\}$. The second is not `minimal', \ie
it does not contain the smallest sets and therefore it is not a
basis for the topology $\mathcal{T}_{s_i}$. Finally, we are left
with $\{(b_1,b_2),(b_3,b_4),(b_5,b_6)\}$, which is the desired
result.

A final note just to mention that the above discussion is liable
to ambiguities that come from the fact that there is not a unique
definition of spatial surface. Instead of the $b$-horizontal as a
surface, we could have taken as spatial surface for example the
subset $\{(c_1,c_2,b_3,b_4,b_5,b_6)\}$, and we would end up with
similar results.

\subsection{Reconstruction of topology---statistical approach}\label{sstatrectop}

As mentioned earlier, different decohering sets of histories may
lead us to different effective topologies. It follows that the
effective topology is a result of our measurements. We could claim
that our system is in a superposition of different effective
`spacetimes'\footnote{Here we still assume that we are in the
non-relativistic regime.} and our choice of measurement causes a
`reduction' to one particular (or to a particular subspace of all
the possible) `spacetime'. In the description above we carry out a
measurement in record space on the `basis' that is related with
spacetime points, \ie events. If this assumption is not satisfied,
the actual choice of our measurements would affect the resulting
topology. It should be pointed out here that this is the generic
case, since we cannot have full knowledge about whether or not our
records capture only configuration space properties and not,
possibly incompatible, momentum space as well. On the other hand,
if our measurements are sufficiently coarse, we could have
compatible `position' and `momentum' measurements.

 Now we are in a position to address how to recover topology
assuming that we can vary slightly one event independently from
the others, and repeat the runs of the experiment. The result of
such variations will be certain changes of the relative
frequencies, that is why we call this process \textbf{statistical}
reconstruction of topology. This procedure fixes the ambiguities
about the `time-slices' that existed due to the non-connected
spatial surfaces, as well as the order of these slices.

\begin{itemize}

\item We have the relative frequencies, $f_j(C_i)$ of each history
$C_i$ with initial condition`$j$'.

 \item We \textbf{vary slightly one event} say event
$p\in\causet$ and repeat the procedure to get the new relative
frequencies of histories $f^p_j(C_i)$. It is important to note
that, provided the variation is small, the set of histories is the
same and only their relative frequencies change. Therefore, all
the considerations that were already made from the mere set of
histories still apply galore.

By observing which histories have changed their frequencies
compared to the undisturbed event case, we can deduce a few
things---for starters, some notion of closeness (or proximity).
The histories whose frequencies are significantly affected by the
perturbation are in some sense `close'.

\item  We consider \emph{each initial condition
separately}\footnote{This to avoid problems related with the
following. Assume that we vary a point $a$ in a way that it has
the same distance with one neighboring point $b$. Then the overall
probability of the $b$ due to symmetry will be invariant, but
depending on which is the initial condition of the system the
probabilities of some histories will increase while other will
decrease with a net probability unchanged. In this way we would
fail to recognize $b$ as a neighbor of $a$.}.For each initial
condition we see the \textbf{probabilities of which histories
alter significantly}.

 \item After we vary the `event', we repeat the experiment exactly
with same initial condition as before and then, by considering the
change in relative frequency of events (not of histories), we can
\textbf{deduce which events are neighbors} (call them
$j$-neighbors, whereby the label `$j$' stands for the initial
condition we consider). We repeat this for all possible initial
conditions. We then consider the union of all these
$j$-neighborhoods to get the total neighborhood of the point we
varied.

In other words we take a small positive number $\epsilon\ll 1$. We
define another function, the difference function, as follows:
\begin{equation}
\delta^p_j: \mathcal{C} \rightarrow [0,1]: \mid
f_j(C_i)-f^p_j(C_i)\mid
\end{equation}

We then consider all the points belonging to the histories
$C_i\in\mathcal{C}$ that

$\delta^p_j(C_i)>\epsilon$. We name them j-neighbors of $p$. So we
have:

\begin{equation}
q\in N^p_j\Longrightarrow \exists \quad q\in
C_i,C_i\in\mathcal{C}\mid \delta^p_j(C_i)>\epsilon\nonumber
\end{equation}

We then consider different initial conditions `$j$' and we group
all the neighbors together to form the neighbors of `$p$' , $N^p$.
\begin{equation}
q\in N^p\Longrightarrow \exists\quad j\mid q\in N^p_j\nonumber
\end{equation}

\item We already know which of these neighbors are (definitely)
not in the same time-slice (:those that both belong to at least
one history), and we can coin them \textbf{`temporal neighbors'}.
Events being in different path-connected components will never
affect each other. Note here that the neighbors that will be
affected, and are definitely not in the same time, are only to the
future of the event in question. Thus, properly speaking, we
should talk about \textbf{`future temporal neighbors'}. With these
in hand, we may get the order of the histories.\footnote{Note that
since we get a direction from the fact that only the `future'
neighbors are affected, helps us recover the order with no doubt
about the overall direction.}

\item Then we mark the events that are neighbors, but not
\emph{temporal} neighbors, as \textbf{`spatial-neighbors'}, and
use them to define proximity in the `time-slice' in focus.

So we define \emph{spatial neighborhood} of `$p$' to be:
\begin{equation}
SN^p\mid q\in[N^p\setminus\cup_i C_i]\quad, \quad p\in
C_i\quad\forall\; i\nonumber
\end{equation}

\item We repeat this procedure varying slightly one by one all the
`events'.

\item From the proximity we deduce the topology of each time slice
in the usual way---\eg as it is done in metric spaces.

\item We will have obtained the topology of each spatial
components. We can then choose an arbitrary partitioning of these
slices to get the total 4-dimensional case. We then check that we
do not have contradiction.This contradiction could be due to, for
example, some event being affected by a change in an event to its
future rather than to its past (:`advanced' and `retarded'
contradiction, respectively).If a contradiction arises, we pick
another `partitioning', so on and so forth, until the correct one
is obtained.

\item By patching all the slices together, we \textbf{recover the
topology} of our `spacetime', or more precisely, of its reticular
substitute .
\end{itemize}

Alternatively, we may consider the closest neighbors to define a
cover of each time-slice, and then find the finitary substitute of
the underlying continuous topology. To find only the closest
neighbors, we need to `tune' the parameter `$\epsilon$' to be
sufficiently big so that it gives only the number of closest
neighbors we want (4 to get a 3-dimensional space in the
triangulations scheme). Along these lines, we first get a
\emph{prebase} from which the topology is unambiguously
reconstructed. We should note here that the above construction
makes `heavy' use of the relative frequencies, not only of the set
of histories. It effectively uses the former to define
neighborhoods.

\section{Toy models}\label{toy models}

\subsection{Double-slit experiment}

We consider a discrete version of the registration screen. This
means that our data will be a discrete distribution of registered
events, each `column' being discretely labelled. We consider the
case that we do not detect which slit the particle passes through,
as well as the case that we do. In both cases we have always the
same initial conditions---a particle is emitted far away from a
barrier bearing two holes. Note also that the particle in question
is assumed not to be a photon, so that it can be detected on the
slit without being absorbed.

\paragraph{Case I: Not detected on the slit.} The particle passes
through the slit. Then it is absorbed by a film so that we can
identify different events by measuring the position of the excited
grain on the film. Note that we need only distinguish the
different events and not their actual position. So, somebody could
have cut the film and glued it back with different order (but the
same for all the repetitions). The correspondence between this
gedanken-experimental scenario and our theoretical scenario above
is the following:

\begin{itemize}

\item single experiment---emitting one particle and registering it

\item the record space---the real line (position-{\it loci} of
registered events)

\item a particular history---an event

\end{itemize}

To recover the configuration space, we (a) assume continuity of
the distribution, and (b) move slightly one point of registration
on the film. By observing the probabilities of the events that are
significantly altered, we define the neighborhood of this `point'
(:proximity neighborhood).

\medskip

\noindent We recover several segments of a line representing the
configuration space. The fact that it is not the whole real line
that is being recovered, is due to the fact that there exist dark
fringes, \ie regions where the particle is never detected.

\paragraph{Case II: Detected on the slit.} The particle passes through the
slit, and a photon, whose frequency depends on which slit the
particle passes through, is emitted. This happens because we have
an oscillator of different frequency on each slit, and when a
particle passes, the oscillator increases its energy level. Then,
upon relaxing back to its ground state, it emits a photon. Then
the particle is absorbed by the discrete screen. The
correspondence between this gedanken-experimental scenario and our
theoretical scenario above is the following:

\begin{itemize}

\item single experiment---emitting one particle, and subsequently
registering it as well as the photon carrying information about
which slit the particle passed.

\item the record space---the real line (position-{\it loci} of
registered events) and the detector of the photon (or the
oscillators). Note here that we can distinguish all events from
each other, but not know anything else about their topological
structure.

\item a particular history---a photon with frequency depending on
which slit the particle passes through, followed by a position on
the discrete screen line.

\end{itemize}

To recover the configuration space, we (a) assume continuity of
the distribution, and (b) we move slightly one point of
registration on the film, or one of the oscillators. As usual, we
recover neighborhoods by small variations of the established
`records'.

\paragraph{What is eventually recovered.} Two points separated from one another, and at a later
time\footnote{The order of the events is recovered, due to the
fact that varying events to the past and only affects the relative
frequencies of the future and NOT visa-versa.} a segment of
straight line (actually a syncopated version of it). Note that the
line does not decay into disjoint segments, as we have no
interference and therefore there are no `dark fringes'.

\subsection{Bath of sensors}\label{sbathsens}

Here we consider a thought experiment that better illustrates the
foregoing ideas. We have a closed box, and in it there are many
(say, $n$) different oscillators, all of different frequency. We
require this to be able to distinguish our `points', but note that
we do not know anything about their structure. We inject a
particle into the box that has the following property: when it is
sufficiently close to one oscillator, the oscillator increases its
energy level. At the final time when we measure things, the
oscillators will relax to their ground state, emitting one photon
of the same frequency as the oscillator that had been excited.

The only other thing we need to assume is that somehow the signals
(photons records) emitted from each oscillator can be
distinguished from those emitted by the same oscillator at a
different time (we need that to have spatio-temporal labels). This
may be done by having, for example, a moving film around the box,
and earlier or later signals from the same oscillator would be
identified by different positions on the film. The fact that each
oscillator may have significantly different `half-life' before it
relaxes, means that signals from different times may be confused.
The important point here is that we only care about the order of
photons coming from the same spatial event, since others can be
distinguished by their different frequency. Finally, we will end
up with a set of different records corresponding to different
events, and all of them will be spatio-temporally distinguishable.

We then infuse the particle into the box, with different momentum
and from different points. Each repetition (with the same initial
condition) gives one possible history. We do it many times for
each initial setup, and we record the results in our data-sheets
(experimental protocols). After this is accomplished, we obtain
the set of possible histories and their relative frequencies. This
is  sufficient, in our non-relativistic case, for deriving the
number of different components each spatial slice has.

We may then vary slightly each oscillator separately, and repeat
the experiment. By this, we will obtain information needed to
recover the topology . In this setting we have the following
correspondences with our theoretical scheme:

\begin{itemize}

\item single experiment---emitting the particle in the box with
some given initial condition, and with a specific setup of the
oscillators, and then record at the final time the photons having
been emitted (maybe read them out directly from the moving film).

\item the record space---photons of different frequency on
different positions on the registering film.

\item a particular history---a collection of photons of different
frequencies (possibly also of different positions on the film, if
a moving film is required for distinguishing events in time).

\end{itemize}

To recover the configuration space, we  do the following:

We assume continuity of the distribution and move slightly one
oscillator at a particular `time'. By noting the probabilities of
which events are significantly altered, we define the neighborhood
of this `point'. Here we need to take into account the union of
all the neighborhoods related to all the possible initial
conditions. We then specify the `temporal' and `spatial'
neighborhoods. We repeat this procedure varying slightly all the
`events' one by one. Using the proximity relation, we deduce the
topology of this slice, as was described in section
\ref{sstatrectop}.

By patching all the slices together, we \textbf{recover the
topology} of our `spacetime'.

\paragraph{What is eventually recovered.} We get the
effective topology of the interior of the box. This includes other
objects that were not known to be there, as well as their time
evolution. So if for example there was a cube of ice in the box
that melted, this will be represented by a cubic hole in the
configuration space that gradually changed shape to become flat.

\section{Conclusions}\label{sconclusions}

Let us summarize what we have done. We have a laboratory in which
we explore a physical system whose configuration space is unknown.
We are able to run the experiments sufficiently many times, either
by leaving the initial conditions unchanged, or by varying them.
We also have another physical system, whose configuration space is
coined {\dff record space}. As a result of each run of the
experiment, the record space acquires a state (a \emph{quantum}
state, in general). In each run, we perform a measurement over the
record space. Which measurement in particular, this is a matter of
our choice.

After multiple runs, we have a set of protocols (data-sheets).
Each protocol tells us which events occurred within a particular
experiment. This set of events is referred to as a history. When
the initial conditions remain unchanged, the arising set of
histories is treated as a decohering set.

Initially, as a result of our observations, we have histories and,
in addition, their relative frequencies. This primary set of
histories we call {\dff fiducial set}.

From the fiducial set, we deduce the number of components of our
`spacetime' (extended configuration space) as well as the number
of components in each `spatial' surface (i.e. moment of time).

We then allow for variations of the records. This yields new
histories which make it possible to deduce proximity on the
fiducial set and hence the topological properties of the
`spacetime'.

As a result, we reconstruct the \emph{effective} topology of the
`spacetime' region involved in our observations. `Effective' means
that we can say nothing about the `true' topology, and that all
our statements are consequences of our observations. The working
definition of configuration space that we employ is the following.
{\dff Configuration space} is the space of all possible
configurations of our system.

\medskip

The topology we recover---the `effective' one---may include holes
and other topological features that result from existing
`potentials' that we do not vary. What could be referred to as the
`true' topology would be something that takes into account only
the background manifold. In this sense it would be like saying
that we may vary not only the initial state of the system in
question, but the initial state of any potential except the
gravitational (which is supposed to account for the `geometry' of
the background manifold).

We emphasize once again that we recover histories
\emph{operationalistically}. The record space is \emph{the only}
source of information we possess about the system we explore. The
effective topology is then regarded as the `best possible' (:as
realistic, or as pragmatic a) picture of the actual configuration
space of the system in focus as one can acquire from her
`experimental intercourse' with it.

Last but not least, some loose, anticipatory connections with the
forthcoming paper \cite{case II} are due here. In the latter, we
develop the {\em relativistic} version of our
`topology-from-inverse histories' theoretical scheme. This
essentially means that, due to a physical upper bound in the
transmission/propagation of (material) signals, one is forced to
focus more on recovering the {\em causal topology} of space{\em
time} from `inverse {\em causal} histories', rather than on just
recovering the topology of `frozen, absolute, fat spatial slices'
(\ie merely of `{\em space}') as we did presently.

\paragraph{Acknowledgments.} We are grateful to Charis
Anastopoulos, Jonathan Halliwell and Serguei Krasnikov, for their
interest in our work and for related feedback, as well as to Chris
Isham, for reading an early draft and making many critical
comments and useful suggestions. The basic ideas underlying this
work were originally conceived shortly prior to and during the
`{\em Glafka-2004: Iconoclastic Approaches to Quantum Gravity}'
meeting in Athens (Greece), generously sponsored by Qualco
Technologies (c/o Dr Orestis Tsakalotos, Qualco's CEO) and by a
European Commission Reintegration Grant (ERG-505432) awarded to
the first author. An outline of those preliminary ideas was
presented briefly at the said meeting by the third author (RRZ).
Subsequently, a major part of the paper was written under the
aegis of the program \emph{T\^ete-\`a-t\^ete in St.Petersburg},
supported by the Euler Mathematical Institute of the Russian
Academy of Sciences. RRZ acknowledges the support from the
research grant No. 04-06-80215a from RFFI (Russian Basic Research
Foundation). PW acknowledges partial funding from the Leventis
Foundation. Finally, IR acknowledges financial support from the EC
via the aforementioned grant.

\end{document}